# GaAs/GaP Superlattice Nanowires for Tailoring Phononic Properties at the Nanoscale: Implications for Thermal Engineering

*Aswathi K. Sivan[1], Begoña Abad[1], Tommaso Albrigi[2], Omer Arif[3], Johannes Trautvetter[1], Alicia Ruiz Caridad[1], Chaitanya Arya[1], Valentina Zannier[3], Lucia Sorba[3], Riccardo Rurali[2], and Ilaria Zardo*[1,4]*

[1]Department of Physics, University of Basel, 4056 Basel, Switzerland

[2]Institut de Ciència de Materials de Barcelona, ICMAB-CSIC, Campus UAB, 08193 Bellaterra, Spain

[3]NEST, Instituto Nanoscienze-CNR and Scuola Normale Superiore, 56127 Pisa, Italy

[4]Swiss Nanoscience Institute, University of Basel, 4056 Basel, Switzerland

ABSTRACT. The possibility to tune the functional properties of nanomaterials is key to their technological applications. Superlattices, *i.e.,* periodic repetitions of two or more materials in different dimensions are being explored for their potential as materials with tailor-made properties. Meanwhile, nanowires offer a myriad of possibilities to engineer systems at the nanoscale, as well as to combine materials which cannot be put together in conventional heterostructures due to the lattice mismatch. In this work we investigate GaAs/GaP superlattices embedded in GaP nanowires and demonstrate the tunability of their phononic and opto-electronic properties by inelastic light scattering experiments corroborated by ab initio calculations. We observe clear modifications in the dispersion relation for both acoustic and optical phonons in the superlattices nanowires. We find that by controlling the superlattice periodicity we can achieve tunability of the phonon frequencies. We also performed wavelength dependent Raman microscopy on GaAs/GaP superlattice nanowires and our results indicate a reduction in the electronic bandgap in the superlattice compared to the bulk counterpart. All our experimental results are rationalized with the help of *ab initio* density functional perturbation theory (DFPT) calculations. This work sheds fresh insights into how material engineering at the nanoscale can tailor phonon dispersion and open pathways for thermal engineering.

**1. Introduction**

In the last few decades, technological miniaturization has led to the creation of smaller and more efficient electronic devices. As devices get smaller, we increasingly face the need for efficient heat management at lower dimensions, typically at the nanoscale. [1,2] Designing materials with tailorable thermal properties is very attractive in this regard. [3,4] The thermal conductivity of materials quantifies the efficiency at which heat energy is transferred within them. Consequently, the capability to tune thermal conductivity becomes essential for effective thermal engineering. Depending on specific applications, there is a necessity to engineer materials with heightened thermal conductivities, such as in high-power electronics[5] and computing devices, as well as materials with reduced thermal conductivities for use in thermoelectric applications[6], thermal barrier coatings[7], etc. In semiconducting and dielectric materials, phonons are the vibrations of the atomic lattice responsible for heat and sound transport. [8] From kinetic theory, the thermal conductivity in semiconductors depend on the phononic properties such as density of states, phonon mean free path, group velocities and phonon energies. [9,10] Consequently, the deliberate manipulation of these phononic properties enables us to create materials with tailor-made thermal conductivity. The advancements in material growth techniques have enabled the growth of artificial material structures with tunable phonon spectra for thermal engineering. [3,11–18] Among the various strategies, the utilization of superlattice (SL) structures emerges as a promising way to tune the thermal property of materials. [11,19,20] SLs are compositionally modulated periodic structures and have been shown to host a plethora of interesting physics. [11,21] Since the first experimental observation of reduced in-plane thermal conductivity in AlAs/GaAs SLs,[22] several theoretical and experimental works have explored SL structures for their potential applications in thermoelectric devices. The theoretical work of P. Hyldgaard and G. D. Mahan,[23] describes a model to calculate the SL thermal conductivity based on modifications to the SL-phonon spectrum. They predicted a reduction in the thermal conductivity of Si/Ge SLs due to the decrease in the group velocity of SL phonon modes. This model was further extended by Tamura *et al*., [24] who calculated phonon group velocities in both Si/Ge and AlAs/GaAs SLs and showed their contributions to thermal conductivity.

SLs with atomically clean interfaces have proven to be ideal systems to study the crossover from particle to wave nature of phonons.[10,25] Indeed, several studies have demonstrated that this crossover can be obtained by controlling the SL periodicity, showing that a distinctive minimum in the lattice thermal conductivity can be achieved.[11,26,27] By taking advantage of wave interference phenomena of phonons, SLs can also serve as an ideal platform for creating phonon bandgap materials by selectively allowing (blocking) phonons with certain energies to propagate through them. Moreover, studies have shown that, we can have control over coherent and incoherent phonon transport in SLs by means of band folding produced by the larger periodicity of the SL and the related changes in the phonon density of states and velocities. [10,11,17,20] The understanding of particle-wave duality of fundamental excitations like electrons and photons have revolutionized modern electronics and optics. By extension, a similar understanding of phonons is highly desirable for designing new ways for manipulating heat flow in materials. However, the need of nanoscale periodicity and atomically smooth interfaces heavily constrains the design of the SL for coherent control of phonon transport.

SL structures can take different forms depending on their constituent layers. Conventional SL structures are formed by the periodic alternation of different materials. [28–30] However, SLs can also be made by periodic alternation of different crystal phases of the same material [31] and by the periodic rotation of the crystal lattice, the so-called twinning SLs [17,32]. The growth of many of these superstructures with atomically abrupt interfaces or minimal interface mixing, however, is limited in thin films. One of the main issues is the lattice mismatch between the constituent layers. Nanowires (NWs) are quasi one-dimensional structures which allow the growth of defect-free axial and radial heterostructures otherwise difficult or even impossible to achieve in planar structures because they can release the strain caused by the lattice mismatch in the radial direction. [33–35] Advancements in epitaxial growth of semiconductor NWs make it possible to grow SL NWs of different types with atomic interface sharpness and minimal interface mixing. This makes SL NWs an ideal platform to envisage and study phonon engineering at the nanoscale for controlled heat transport. In recent years, there has been an increase in both the theoretical and experimental studies of SL NWs with a focus on their thermal properties.[17,36–38] Most of the works done in NW SLs from the perspective of phonon engineering explores twinning or crystal phase SLs. In this work we study the phononic properties of SL NWs made of two different constituent materials: GaAs and GaP. Combination of materials with different elastic constants creates an enhancement of phonon wave interference effects such as phonon bandgaps.[16] Both GaAs and GaP are highly relevant materials in the realm of photonic and opto-electronic devices with applications in photovoltaics[39–41], light emitting devices,[42] transistors,[43] etc. The possibility to combine these two materials at the nanoscale will create a versatile material system for photonic and phononic applications. However, the lattice mismatch between GaAs and GaP is approximately 3.7%, posing growth constraints for planar superlattices.[21] The NW geometry enables the combination of GaP and GaAs epitaxially without formation of misfit dislocations at the interface.

In this work, we investigate the phononic properties of GaAs/GaP SL NWs with periodicities ranging from 4.8 nm to 10 nm. We present the results of inelastic light scattering experiments, such as Raman microscopy and Brillouin light scattering interferometry of these SL NWs, and we show that the phononic properties of GaAs/GaP SL NWs can be tuned by controlling the SL periodicity. We demonstrate that as the SL periodicity increases, the number of phonon modes increases. Our experimental results together with the ab initio calculations indicate a reduction in the electronic bandgap in the SL structure compared to the bulk materials. Our study show that NWs can serve as a template for combining different materials at the nanoscale for designing new material systems with tunable phononic and optoelectronic properties.

## 2. Results and Discussion

GaAs/GaP SL NWs with different periodicities are grown using Au-assisted chemical beam epitaxy (CBE) on GaAs (111) B substrates. The GaAs/GaP SL NW growth process is described in detail elsewhere.[44] Figure 1 (a) illustrates the four step Au-catalyzed epitaxial growth process of a typical GaAs/GaP SL NW using CBE. Figure 1(b) shows the Scanning Electron Microscopy (SEM) image of a representative as-grown sample of GaAs/GaP SL NWs with relatively homogeneous lengths (3.58 ± 0.24 μm) and diameters (40 ± 5 nm). The structural and chemical composition of the GaAs/GaP SL NWs were studied using a transmission electron microscope (TEM) and energy-dispersive X-ray analysis (EDX) measurements. The period L of the SL structure is the sum of the lengths of consecutive GaAs ($L_{GaAs}$) and GaP ($L_{GaP}$) segments, *i.e.*, L= $L_{GaAs}$ + $L_{GaP}$. Figure 1(c) shows the TEM image of a SL NW with a nominal period of 10 nm ($L_{GaAs}$ = 5 nm and $L_{GaP}$ = 5 nm). The detailed high resolution TEM (HR-TEM) image of a SL NW with 4.8 nm long period is found in SI in Figure S1. The TEM image shows alternatingly spaced GaAs and GaP segments along the growth direction of the NW and from the HR-TEM in Figure S1, we measure L as 5.3 ± 0.6 nm, i.e. consistent with the nominal value, using DigitalMicrograph$^{TM}$. Figure 1(d) shows the EDX elemental mapping of a whole NW, in 4 different positions along the growth axis, from the base (bottom panel) to the tip (upper panel). The EDX elements are represented in false colors as follows: As in green, Ga in blue and P in red. The NW is mapped from bottom to top as follows: at the bottom of the wire, we have a first segment composed of GaAs, this is shown in the EDX map in the color combination of blue and green as cyan in Figure 1(d). After this segment, we have a long GaP segment, shown in the color combination of red and blue as magenta, which is followed by the SL segment, depicted in Figure 1 (d) as a combination of red, blue and green. The SL segment consists of approximately 100 repetitions of the GaAs + GaP unit. At the top part of the NW, we have a final section of GaP segment shown again in the false color combination of red and blue in Figure 1(d). The other samples of this work have the same design and compositional sequence, they differ only for the nominal SL periodicity: L = 6.0 nm ($L_{GaAs}$ = 3.0 nm and $L_{GaP}$ = 3.0 nm) and L = 10.0 ($L_{GaAs}$ = 5.0 nm and $L_{GaP}$ = 5.0 nm). A reference sample of NWs without SL, i.e made of a 0.5 μm long GaAs stem followed by 3 μm of GaP segment, was also grown.

In order to probe the phonon spectra of GaAs/GaP SL NWs, we performed inelastic light scattering experiments such as Raman microscopy on single GaAs/GaP SL NWs with different period lengths and Brillouin light scattering interferometry on the as-grown NW ensemble.

In Figure 2(a), we display the experimental Raman spectrum obtained from the GaP reference NW in the $\bar{x}(z,z)x$ scattering configuration in Porto notation, as described in the Methods section. The experimental data are plotted with dark green spheres and the solid lines corresponds to the Lorentzian fitting. The red solid curve corresponds to the cumulative peak fitting, while the purple curves correspond to deconvoluted Lorentzian used to fit the individual peaks. Between 340 cm$^{-1}$ and 410 cm$^{-1}$, the experimental data can be fitted with 4 peaks at frequencies ~352.2 cm$^{-1}$, ~362.2 cm$^{-1}$, ~371.0 cm$^{-1}$, and 400.7 cm$^{-1}$, which can be assigned to the $E_2^H$ mode, the $A_1/E_1$ transversal optical (TO) mode, the surface optical (SO) mode, and the longitudinal optical (LO) mode, respectively. [45–47] The calculated spectrum of bulk wurtzite (WZ) GaP for the configuration $\bar{x}(z,z)x$ is shown in Figure 2(b). We observe a peak at ~ 362.6 cm$^{-1}$ which corresponds to the TO mode of GaP in good agreement with the experimental observation. We observe a low intensity LO mode in the experimental spectrum, in principle forbidden and thus absent in the calculated spectrum. This observation is due to the usage of a high numerical aperture (NA) objective and due to the partial relaxation of the selection rules, similar to what is observed and explained in De Luca *et al.* [17]. Similarly, we observe a peak around the frequency of the $E_2^H$ mode, a signature of the WZ crystal phase. The $E_2^H$ mode is also forbidden in the $\bar{x}(z,z)x$ configuration and its observation can also be assigned to the use of a high NA objective as well as due to the small deviations in the selection rules. It is worth noticing that, it has been shown that the $E_2^H$ has a strong wavelength dependence . [48] Also, we cannot exclude that the signal we observe is due to an asymmetric broadening of the GaP TO mode because of the anharmonicity of the material.[49] In Figure S2, we have presented the results of the Raman spectroscopy done in the $\bar{x}(y,y)x$ configuration for the reference NW where the presence of the $E_2^H$, which is allowed in this scattering geometry, can be better appreciated.[17,50,51] The SO, on the other hand, is due to finite size effects because of the NW geometry,[45,52] and its frequency depends mainly on the dielectric constant of the surrounding medium as well as on the diameter of the NW. Therefore, it is absent in the computed spectrum, as the calculations were performed in a bulk system. In Figure 2 (a), we also see a small peak at around 300 cm$^{-1}$ arising from the Si substrate beneath.[53] Figure 2 (c) is a schematic of the reference NW from which the spectrum in Figure 2 (a) was obtained.

In Figure 2(d), we display the spectrum obtained experimentally from a 4.8 nm period GaAs/GaP SL NW (whose schematic is depicted in Figure 2(f)) in the $\bar{x}(z,z)x$ configuration, probing the region of the NW embedding the SL. In the spectral region of the phonon modes of GaP we observe several modes between 340 cm$^{-1}$ and 400 cm$^{-1}$. The most intense peaks in this spectral range are at frequencies 358.3 cm$^{-1}$, 352.8 cm$^{-1}$, and 388.2 cm$^{-1}$ in the order of decreasing intensities. These new peaks appearing are assigned to modes arising from the SL. We can also see several Raman peaks in the spectral region (between 240 cm$^{-1}$ and 300 cm$^{-1}$) close to the phonon modes of GaAs, the most intense peak being around 276.4 cm$^{-1}$. The other peaks are located at 261.3 cm$^{-1}$ and 287.6 cm$^{-1}$. In Figure 2 (e), we present the calculated spectrum for a GaAs/GaP SL with a period of 5.12 nm. In the calculated spectrum, we have fixed the full width at half maximum (FWHM) of the peaks at 5 cm$^{-1}$, since it is similar to the experimentally measured peak broadening. The theoretical spectrum of the SL structure contains several peaks between 240 cm$^{-1}$ and 300 cm$^{-1}$ as well as between 340 cm$^{-1}$ and 400 cm$^{-1}$ (not distinguishable in the convoluted spectrum with the given FWHMs; a list of the computed Γ-point frequencies, setting a threshold intensity indicative of their experimental detectivity, is given in Table S1). Experimental and theoretical spectra obtained from a SL with longer period (10 nm and 6.39 nm, respectively) are shown in Figure S3 of the supporting information.

These several phonon modes can be understood in terms of the back folding of the phonon dispersion of the constituent materials of the SL or, equivalently, from the fact that the SL period dictates a much larger unit cell, containing many more atoms than the WZ primitive cells of the constituent materials.[54] The phonon dispersion of the GaP/GaAs SL with a period of 6.39 nm obtained through ab initio calculations can be found in Figure S4 of the supporting information. There, the increased number of phonon modes at the Γ-point can be appreciated, though it is worth noticing that not all of them are Raman active (see also Table S1).

Furthermore, to prove the origin of the phonon modes in the two different frequency ranges (*i.e.* 240–300 and 340–400 cm$^{-1}$), we then analyzed the corresponding eigen-displacements. Figures S5(a) and S5(b) in the SI show the schematic for two phonon modes at 288.1 cm$^{-1}$ and 367.2 cm$^{-1}$, respectively, selected from Table S1. In the schematic, the blue, yellow and green spheres represent Ga, As and P atoms, respectively. The phonon mode at 288.1 cm$^{-1}$ involves vibrations of Ga and As atoms while the phonon mode at 367.2 cm$^{-1}$ involves vibrations of the Ga and P atoms. Therefore, for sake of simplicity, from now on, we will refer to SL modes in the low frequency range as to GaAs-like SL mode and to more in the high frequency range as to GaP-like SL modes. For the sake of generality, in Figure S6 of the supporting information we also show the eigen-displacement of a Raman inactive, low frequency mode, whose vibrations involve atoms on both the GaP and the GaAs region.

We also performed spatially resolved one-dimensional μ-Raman measurements along the length of the GaAs/GaP SL NW of period 4.8 nm as well as that of the GaP reference NW studied in Figure 2. We collected Raman spectra every 300 nm since the spatial resolution is also limited by our spot size, our spot size is around 420 nm. In Figure 3 (a) the result of the Raman scan of the GaP reference NW is plotted in a 2D false colormap with x-axis showing the Raman shift and the *y*-axis showing the laser position along the NW. From the map we see only signal arising from GaP, *i.e.* the TO mode at ~ 362 cm$^{-1}$. In Figure 3 (b), we present the results of a similar scan performed on the GaAs/GaP SL NW of period 4.8 nm shown in Figure 2(d). In the map we see the signal from GaP, followed by SL signal for about 1 μm, and a weak GaP signal after the SL segment. We do not see any signal arising from the GaAs segment in the reference NW or the SL NW. This could possibly be due to breaking of the NW at the junction between GaAs and GaP during the transfer using a micromanipulator. In Figure 3(c) we present individual spectra extracted from Figure 3(b). The black dotted line is a guide for the eye to observe the downward shift in the frequency of the high intensity peak from the GaP-like modes. In Figure 3(d), the shift of the highest intense GaP-like SL peak from the TO of WZ GaP measured in the reference NW is shown as a function of the position along the NW. Outside the SL, the frequency of the highest intensity peak in the GaP-like modes coincides with that of the WZ GaP TO mode. However, on entering the SL segment, the frequency of the highest intensity down-shifts considerably. This is a further indication of the different origin of these modes: the SL acts like a different material system with its own characteristic phonon modes.

After observing the appearance of the SL modes in the 4.8 nm period GaAs/GaP SL NW, we tested the tunability of the phonon modes by varying the SL period length. In Table 1, we present the calculated Raman active phonon modes at the Γ point for three different periods, the threshold intensities are set according to their experimental detectability like in Table S1. From Table 1, we see that as period increases, the number of phonon modes also increases. Similarly, we also see that different periods have distinct phonon modes. We have performed both Brillouin and Raman scattering experiments on NWs with different SL periods to understand the effect of SL periodicity on the acoustic and optical phonons, respectively.

The acoustic phonon modes were probed by Brillouin light scattering (BLS) interferometry.[55,56] Figure 4(a) shows the BLS spectra of SL NWs with different periods as well as that of reference NW. The BLS spectra were fitted with Lorentzian curves to extract the peak position of the signal. From the reference sample, we fit 4 peaks, centered around 25.6 GHz, 30.8 GHz, 37.1 GHz, and 101.3 GHz, while for a SL with L= 10 nm we extracted 8 peaks centered around 25.4 GHz, 29.2 GHz, 36.1 GHz, 48.3 GHz, 68.1 GHz, 90.7 GHz, 95.43 GHz, and 99 GHz. For the SL sample with SL period 6 nm we fit the spectrum with 5 curves centered around, 26.4 GHz, 36.6 GHz, 49.6 GHz, 72.1 GHz, and 95.6 GHz. And for a SL with L= 4.8 nm, we fit the data with 7 peaks centered around 26.4 GHz, 36.1 GHz, 40.5 GHz, 50.1 GHz, 68.5 GHz, 86.1 GHz, and 97 GHz. From our experimental results, we observe a greater number of phonon modes in the SL NWs compared to the reference NW. To understand the modifications in the acoustic phonons in SL structure, we computed phonon dispersion of WZ GaP as well as GaAs/GaP SLs of different periods. In figures 5(b) and 5(c), we show the computed band structure of WZ GaP and a GaAs/GaP SL with L= 1.3 nm, respectively. We can see that compared to the GaP bands, the 1.3 nm long period has a greater number of phonon bands in the dispersion. Dispersions of a SL with L=6.39 nm is shown in Figure S4. In our BLS experiments, we use a laser line at 532 nm for the excitation, the wave vector is given by $q = 4\pi n/\lambda$ nm$^{-1}$, where n is the refractive index[57]. In our case, we get the q value at around 85 μm$^{-1}$. From the calculations done for bulk GaAs/GaP SL with periodicity 6.39 nm, within our experimental detection window (<120 GHz), there are two modes at around 35 GHz and around 75 GHz. In our experiments, for the GaAs/GaP SL NW with L=6 nm, we see peaks around 36 GHz and 71 GHz, in very good agreement with the calculation. Within our experimental detection window, i.e. for phonon frequencies between -120 GHz and 120 GHz, for q values close to 85 μm$^{-1}$, we obtain a good agreement with calculation. However, we also observe a few phonon modes which do not appear in the calculations. This deviation between calculation and experiment could be explained as due to a combination of several reasons, the first is the difference between experimentally measured and calculated phonon frequencies leading to calculated modes falling out of the experimental detection window. In our calculations of the dispersion relationship, we do not take into account the surface phonon modes arising from surface ripple mechanism,[58,59] which is often observed in Brillouin interferometry of metals and semiconductors. Our bulk calculations also do not take into account confined acoustic phonon modes reported for e.g. by Kargar and coworkers in GaAs NWs with diameters in 100-150 nm range.[57]. We have tabulated the calculated acoustic phonon modes with frequencies less than 1 THz for a SL with periodicity of 6.3 nm in the SI in table S2.

Figure 5 presents the overview of the results on tunability of the optical SL phonon modes as a function of SL periodicity for GaAs/GaP SL NWs. In Figure 5 (a) the calculated Raman spectra obtained for different SL periodicities (3.83 - 6.39 nm) are shown as solid lines, while in Figure 5(b) we present the μ-Raman spectra taken from SLs of different periodicities (4.8 - 10 nm period). The experimental data are represented by colored spheres while the solid curves show the Lorentz curves used for fitting. The quantitative analysis of the SL modes is summarized in Figure 5(c). Open circles represent the peak positions extracted from calculated spectra, while filled circles represent the peak position extracted from Lorentzian fits of the experimental Raman measurements. The colors yellow and green are used to denote GaAs- and GaP-like SL modes, respectively. We also observe, both in the experimental and in the calculated spectra, that as the periodicity increases, the number of Raman active phonon modes also increases, as it is expected due to the increased number of atoms in the unit cell of the periodic superstructure and the related backfolding of the phonon dispersion. One major difference between the calculated and experimental spectra are the relative intensity between the GaAs-like and GaP-like modes. For calculated spectra, the intensity of GaAs-like modes is higher than GaP-like modes, while for experiments performed using a 488 nm laser, the GaP-like modes are more intense. This is possibly due to resonant conditions. Namely, the SL exhibits also a modified electronic band structure, whose details (e.g. number of bands, their dispersion, and energy gaps) depend on the periodicity of the SL. Therefore, we performed wavelength dependent μ-Raman measurements on a GaAs/GaP SL NW with L= 4.8 nm to test the effects of the excitation wavelength on the SL phonon modes. The results of wavelength dependent measurements are summarized in Figure 6. In our wavelength dependent Raman studies, we use $CaF_2$ as the reference material as it has a constant Raman tensor in the range 1.8–3.8 eV. [60,61]. We first normalized all the measured spectra with the intensity of the $CaF_2$. In Figure 6(a), we present the normalized results of Raman measurements from a SL NW with L = 4.8 nm excited with 3 different wavelengths. The data in violet, green, orange, and red curves represent spectra obtained with excitation wavelengths 488 nm, 514 nm, 561 nm, and 633 nm, respectively. One striking observation as we increase the excitation wavelength is that the relative contribution of GaAs-like modes compared to GaP-like modes increases. This is clearer in Figure 6(b), shown as the ratios of intensities of the most intense peak for the GaP-like mode (at around 356 cm$^{-1}$) to the intensity of the most intense

GaAs-like mode (at around 276 cm$^{-1}$). We see that the value of this ratio goes from around 3 to less than 0.5 as we increase the excitation wavelength from 488 nm to 561 nm, this value then slightly increases to 0.7 at 633 nm. In Figure 6(c) we plot the intensities of the GaAs-like and GaP-like modes the violet, green, orange, and red symbols represent excitation wavelengths 488 nm, 514 nm, 561 nm, and 633 nm, respectively. For the SL NW excited with different wavelengths, we see that for GaAs-like phonon mode around 276 cm$^{-1}$ shows a significant increase in signal at 561 nm and 633 nm. For GaP-like phonon modes, the phonon mode around 356 cm$^{-1}$ shows comparable intensities for 633 nm, 561 nm and 488 nm and a lower intensity for 514 nm. All other SL phonon modes presented comparable intensities at different excitation wavelengths. In Figure 6(d), we present results of similar measurements done on the reference NW. For the WZ GaP signal of the reference NW, we see that the signal intensity decreases for increasing excitation wavelength, with the highest intensity for 488 nm (2.54 eV), followed by 514 nm (2.41 eV), then by 561 nm (2.21 eV) and lowest for 633 nm (1.95 eV). This trend in intensities point to approaching resonant conditions as we increase excitation energy. It is worth to point to a previous work done using resonant Raman studies to extract electronic band structure of WZ GaP NW bandgap have shown that E1(LO) mode shows resonances at 2.38 and 2.67 eV, and A1(LO) exhibits resonance at 2.67 eV.[62] The different response of the SL and reference NW to different excitation wavelengths also is a strong indication of the modification of the electronic band structure in a SL as compared to the constituent materials. In order to further understand effect of SL periodicity on the electronic band structure, we performed Density Functional Theory (DFT) calculations with the projector augmented wave method [63] and the Local Density Approximation (LDA) of three GaAs/GaP SLs with different periods. Local and semi-local approximation to the exchange-correlation energy, such as the LDA, are known to underestimate electronic bandgaps. However, in a previous work, Giorgi *et al.* [64] performed $G_0W_0$ calculations, which provide bandgaps in good agreement with the experiments, obtaining a correction of the DFT-LDA bandgap of WZ bulk GaAs and GaP of ~1 eV (0.91 and 0.96 eV). Therefore, as a first approximation, we assume a similar correction for the bandgap of GaAs/GaP SLs as well. The results of our calculations along with $G_0W_0$ corrections are tabulated in Table 2. From the calculation, we find that the bandgaps of the GaAs/GaP SLs are lower than that of the bulk WZ GaAs and GaP, indicating a type II band offset. Our calculations also indicate that the bandgap decreases with increasing period. The different excitation wavelength dependence of the Raman response of the GaP reference sample and the GaAs/GaP SLs can thus be attributed to the different electronic band structures.

## 3. Conclusions

In this work, we reported the phononic properties of GaAs/GaP SL NWs with different periods using inelastic light scattering experiments. Our results are corroborated with ab initio theoretical calculations. We were able to assign the phonon modes stemming from the SL, associating their frequencies to the atomic species involved in the vibration. We also showed the tunability of the phononic spectrum studying the dependence of both acoustic and optical phonon modes as a function of the SL period. Namely, the number of phonon modes increases with the SL period, which results from an increased number of atoms per unit cell. The investigation of the Raman response to different excitation wavelengths of SL compared to the reference system hints at a different electronic band-structure, which is also expected to be SL period dependent. Hence, we demonstrated the possibility to obtain metamaterials with designed functional properties by controlling the SL period. Our results show that NW SLs allow to create materials with different phonon dispersion which in turn has implications in designing thermoelectric materials with reduced thermal conductivity. Our results also have a significant impact on creating designed materials with tunable thermal, acoustic, and optoelectronic properties.

## 3. Methods

Growth: The GaAs/GaP SL NWs were grown on GaAs (111)B substrates by Au-assisted Chemical Beam Epitaxy (CBE) in a Riber Compact-21 system. The metalorganic (MO) precursors used for the NW growth were triethylgallium (TEGa), tertiarybutylarsine (TBAs), and tertiarybutylphosphine (TBP). The GaAs (111)B substrates were first coated with a 0.1 nm thick Au film at room temperature in a thermal evaporator and then transferred to the CBE growth chamber. Prior to start the growth, the samples were annealed at 520 ± 5 °C under TBAs flux for 20 min in order to dewet the Au film into nanoparticles and to remove the surface oxide from the GaAs substrate. The NW growth protocol consisted of four steps. In step I, the growth of a 0.5 μm long GaAs stem was performed for 60 min at 510 ± 5 °C by using MO line pressures of 0.7 and 1 Torr for TEGa and TBAs, respectively. Afterward, the temperature was ramped up to 560 ± 5 °C in 5 min to initiate the growth of the GaP segment (step II) with a direct switch of the precursor fluxes, without any growth interruption. The GaP segment was grown using MO line pressures of 0.7 and 2 Torr for TEGa and TBP, respectively. In step III, the growth of GaAs and GaP alternating segments forming the SL was performed at 560 ± 5 °C by using the same MO line pressures. Finally, (step IV), a 1 μm long GaP top segment was grown by keeping unchanged the growth parameters (growth temperature and MO fluxes). The growth was terminated by switching off the TEGa flux and cooling the sample under TBP flux.

TEM / EDX: NW transparent to the electron beam were dispersed carbon holey grid. HR-TEM and EDX measurements were performed using a JEOL JEM F220 microscope operated at 200 kV and equipped with an EDX spectrometer.

Raman: For the µ-Raman experiments, we transferred single NWs onto a Si substrate with 525 nm of silicon nitride coating using a micromanipulator. The polarization resolved Raman measurements were performed in back scattering geometry with the help of a Horiba T64000 triple spectrometer in subtractive mode with a 1.800 g/mm grating and a liquid nitrogen-cooled CCD detector. Whenever not specified otherwise, the presented spectra were collected using a 488 nm excitation wavelength. We used a power below 120 µW to perform the Raman experiments. At room temperature, we used a 100x objective with a high NA of 0.95 for focusing the excitation laser and for collecting the scattered light. We use the conventional Porto notation of the form $k_i(\varepsilon_i,\varepsilon_s)k_s$ to indicate the polarization configuration of our Raman measurements, where $k_i$, $\varepsilon_i$, $\varepsilon_s$, and $k_s$ are direction of propagation of incident photon, direction of polarization of incident photon, direction of polarization of scattered photon, and direction of propagation of scattered photon, respectively. In this work, since we use back scattering geometry, we assume the incident and scattered photon wave vector to be antiparallel and parallel to the $x$ axis. Therefore, the polarization vectors will lie in the plane perpendicular to the directions of propagation, *i.e.*, the $yz$ plane. In our experiments, we consider the NW growth axis to lie along the $z$ axis. For example, the configuration $\bar{x}(z,z)x$, which is the main scattering geometry used in this work, represents a measurement with excitation and detection polarization along the NW growth axis. For the sake of comparison, we have used a reference NW sample consisting of a 500 nm long GaAs segment followed by 3 µm of GaP stem.

Brillouin: We used a 6-pass tandem Fabry Perrot interferometer, to investigate the low frequency acoustic phonons. [55,56,65] The measurements were performed on as-grown NWs on GaAs substrate using a 532 nm laser at room temperature. We used a 100x objective to excite and collect the scattered light for perform the BLS measurements.

DFPT calculations: The Raman susceptibility tensors were calculated within density-functional perturbation theory (DFPT)[66] with the ABINIT code [67,68] by computing the third derivative of the total energy (twice with respect to the application of an electric field, i.e., incident and scattered light polarization vectors, and once with respect to the phonon displacement coordinates [69]). We used the LDA for the exchange-correlation energy functional and norm-conserving pseudopotentials. We used a plane wave cutoff of 37 Ha, an energy cutoff for the fine fast Fourier transform of 76 Ha, and a strict convergence criterion of the wave function residual norm of 10-22. We studied SLs with periods of 1.28, 2.56, 3.83, 5.12, and 6,39 nm, constructed by piling up unit cells of wurtzite GaAs and GaP along the [0001] crystal axis. The Brillouin zone was sampled with a grid of $8 \times 8 \times N$ k-points, decreasing N from 3 to 1 when going from the shorter period to the longer period. Prior to the calculation of the Raman spectra all systems were structurally relaxed, optimizing both the atomic positions and the cell lattice vectors, without applying any additional constraint. The calculations have been performed in a bulk system. Further details can be found e.g. in Ref. [17].

FIGURES

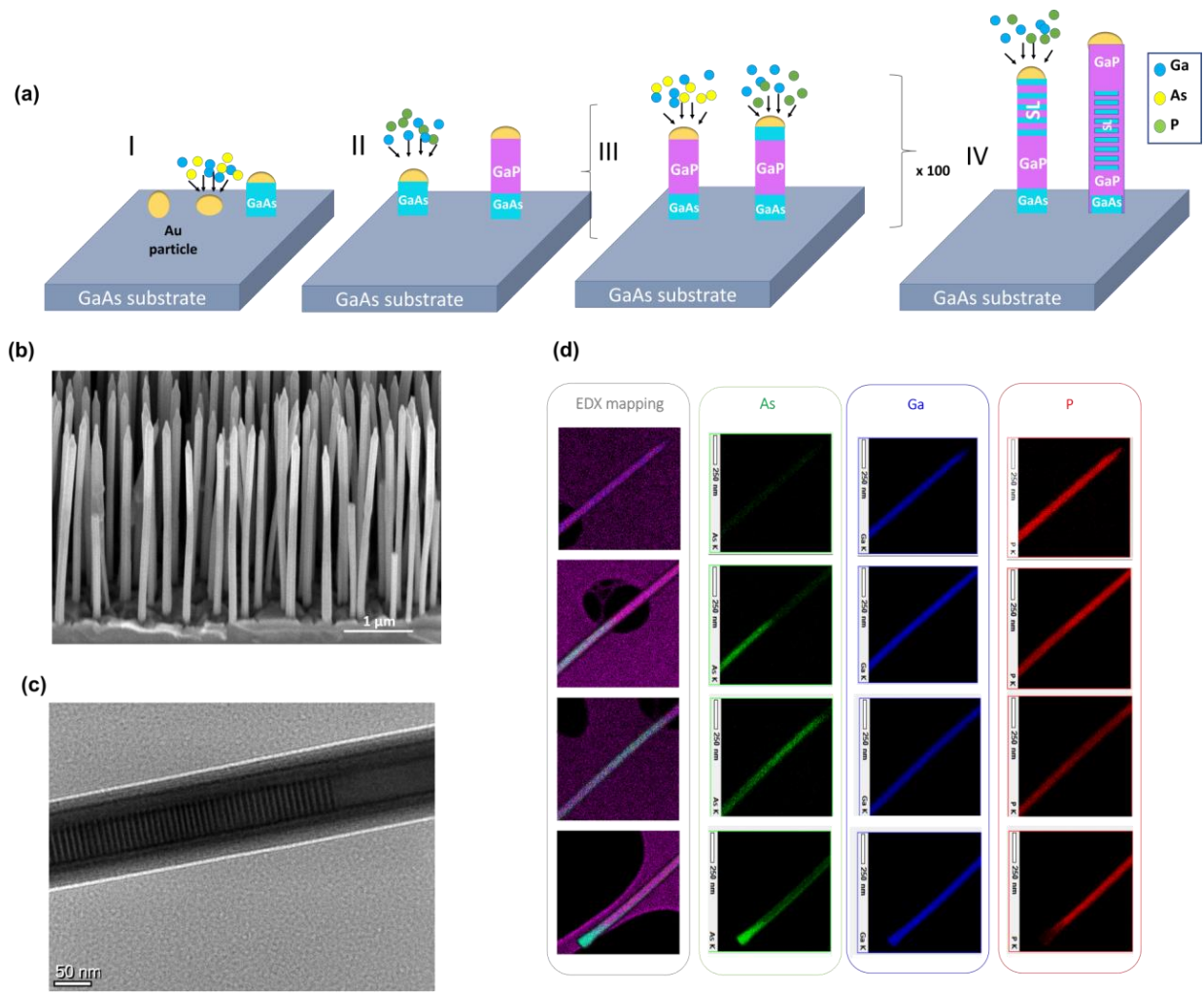

Figure 1. GaAs/GaP superlattice nanowires. (a) Schematic illustration of the four-step growth process of a typical GaAs/GaP SL NW using Chemical beam epitaxy; (b) SEM image of as-grown free standing GaAs/GaP SL NWs grown on GaAs substrate; (c) TEM image of a typical SL NW of a nominal period 4.8 nm dispersed in a holey carbon grid investigated in this work; EDX map in false colors of a NW taken from the same sample as the one shown in (b); (d) the EDX mapping in false colors shows the distribution of elements along the length of the wire showing the presence of the GaAs/GaP SL segment. The elemental maps of one NW are represented as follows: green for As, blue for Ga and red for P.

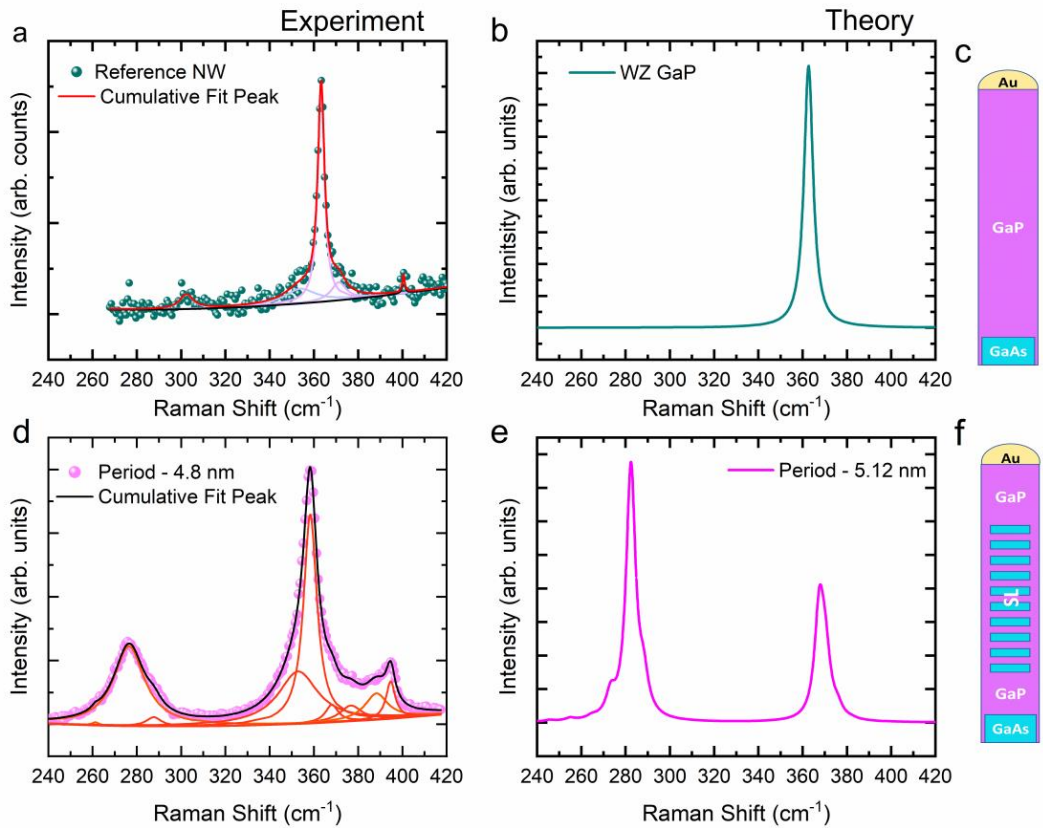

**Figure 2**. (a) Measured spectrum of GaP from reference NW with the experimental data plotted in green spheres, the cumulative Lorentzian fit is shown in solid red line, the deconvoluted Lorentzian fits are shown in violet curves, and the baseline in black; (b) Calculated spectrum of bulk wurtzite GaP; (c) A schematic of the GaP reference NW from which the spectrum in (a) is obtained; (d) Measured spectrum of a GaAs/GaP SL NW with a period of 4.8 nm with the experimental data shown in magenta spheres and the cumulative Lorentzian fitting in solid black curve and the red curves show the deconvoluted Lorentzian fits, and orange line shows the baseline; (e) The calculated Raman spectrum of a GaAs/GaP SL NW with a period of 5.12 nm; (f) A schematic of the GaAs/GaP SL NW from which the spectrum in (d) is taken.

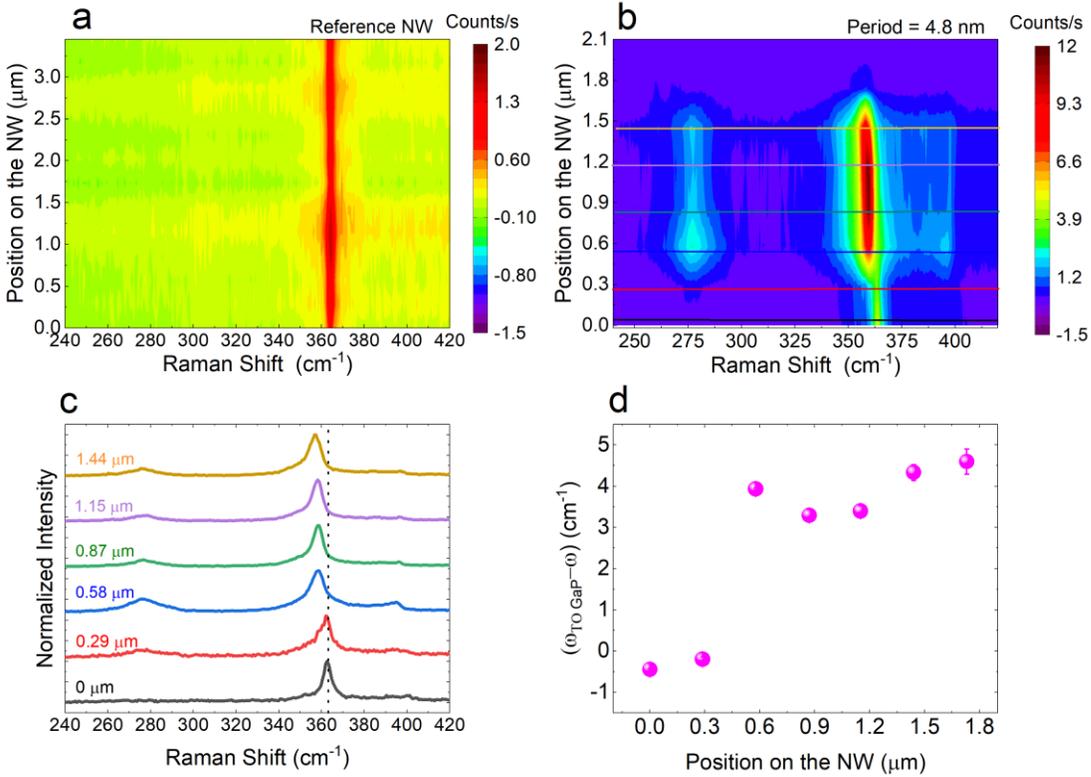

**Figure 3.** Spatially resolved Raman map of (a) GaP reference NW; (b) GaAs/GaP SL NW of period 4.8 nm; (c) Individual spectra corresponding to the horizontal lines extracted from (b); (d) Relative shift of the mode highest intensity mode in the GaP-like region with respect to the frequency of the experimental GaP TO mode as a function of the position along the NW.

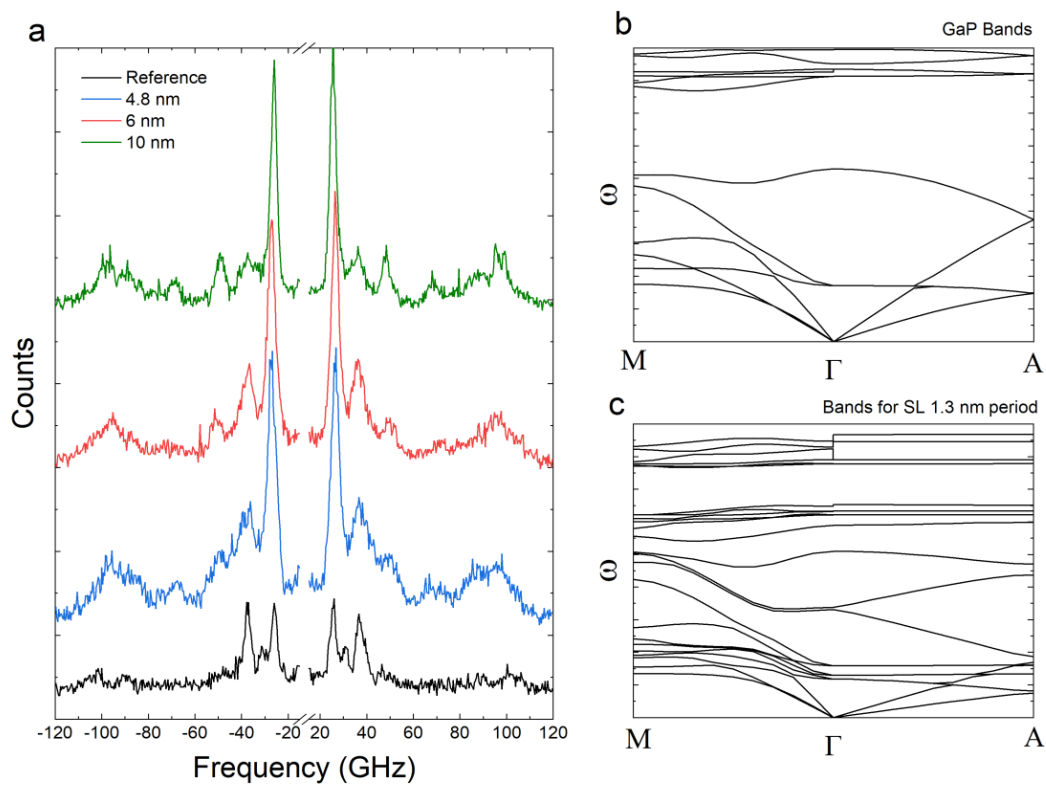

**Figure 4.** (a) Brillouin scattering spectrum of GaAs/GaP SL NWs of different periods, the curve in black represents the signal from the GaP reference, while the curves in blue, red, and green are collected from 4.8 nm, 6 nm and 10 nm periods SLs, respectively; (b) Calculated band structure of bulk WZ GaP; (c) Calculated band structure of a SL of 1.3 nm period.

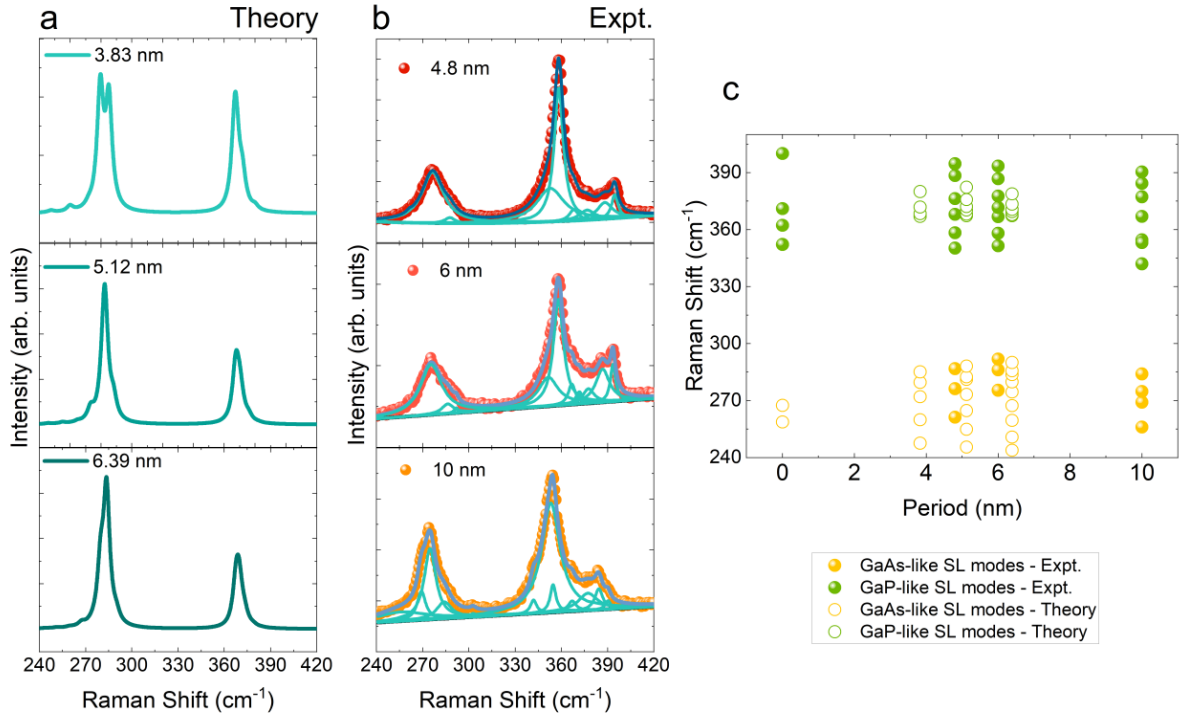

**Figure 5.** (a) Calculated Raman spectra of 3 different SL periods; (b) Experimental Raman spectra obtained from 3 different SL period along with the deconvoluted Lorentz fitting; (c) the Raman peak positions extracted from both experimental (filled circles) and calculated (open circles) Raman spectra as a function of period length.

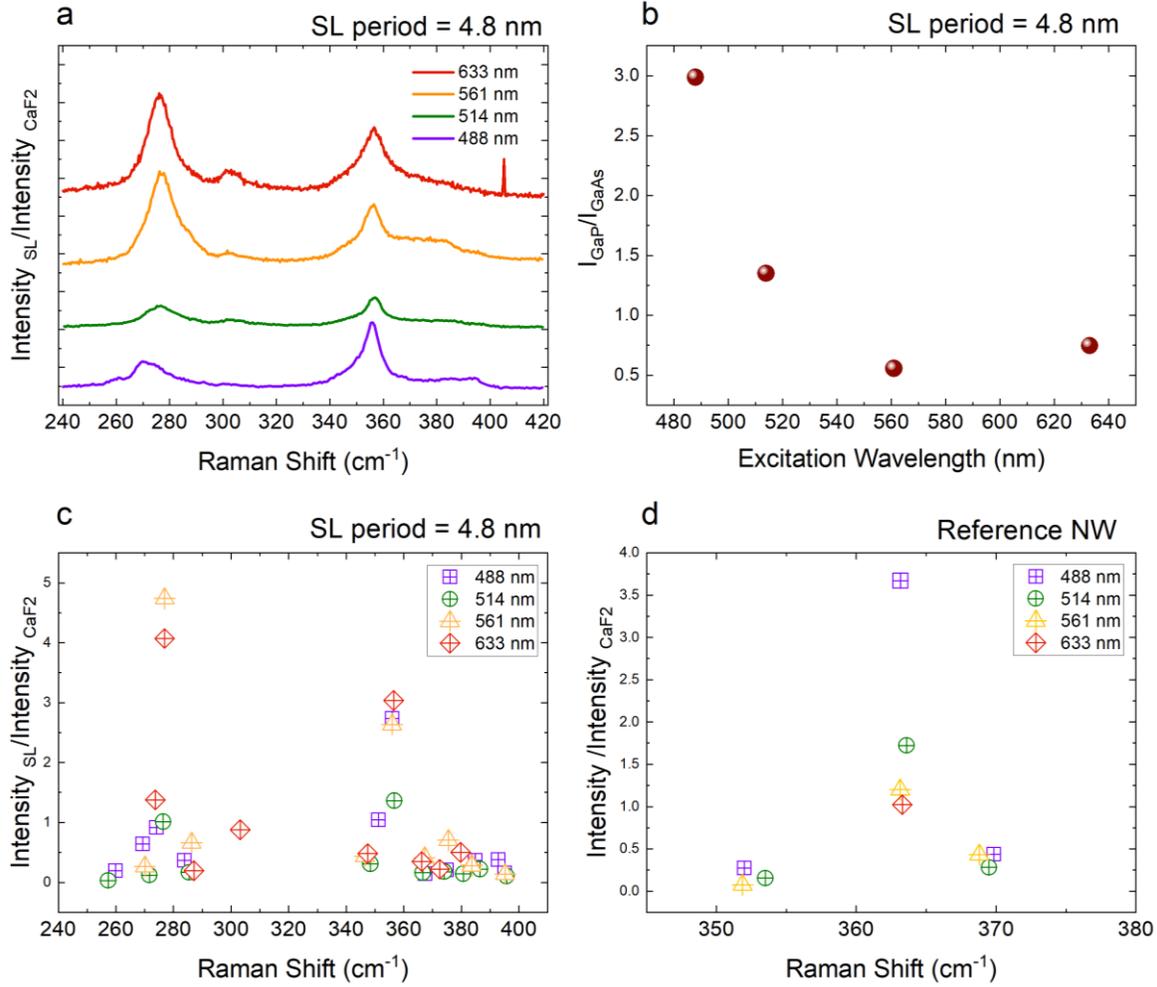

**Figure 6.** (a) The y-offset plots of Raman spectra of a 4.8 nm period GaAs/GaP SL NW obtained with 488 nm (violet), 514 nm (green), 561 nm (orange), and 633 nm (red). The individual spectra are normalized using $CaF_2$ Raman intensity for the given excitation wavelength; (b) The ratio of intensity of GaP-like mode around 356 $cm^{-1}$ to intensity of GaAs-like mode around 276 $cm^{-1}$ as a function of excitation wavelength; (c) The intensities (normalized with $CaF_2$ intensity) of different phonon modes as a function of the Raman shift extracted from (a) with same color scheme as in (a); (d) The intensities (normalized with $CaF_2$ intensity) of different phonon modes as a function of the Raman shift for a reference NW, with same color scheme as in (a).

TABLES.

**Table 1.** The calculated frequencies of Raman active phonon modes for a GaAs/GaP SL structure with 3 different periods. Phonon modes with a Raman intensity more than 100 times smaller than the one of the most intense mode were considered non-detectable.

| Frequencies for L=3.83 nm (cm$^{-1}$) | Frequencies for L=5.12 nm (cm$^{-1}$) | Frequencies for L=6.39 nm (cm$^{-1}$) |
|---|---|---|
| 247.53 | 245.55 | 243.87 |
| 259.88 | 254.87 | 250.83 |
| 272.02 | 264.73 | 259.57 |
| 279.57 | 273.24 | 267.39 |
| 285.13 | 281.13 | 274.60 |
| 366.94 | 282.45 | 279.78 |
| 368.35 | 288.12 | 283.64 |
| 371.79 | 367.25 | 285.71 |
| 379.98 | 368.09 | 290.04 |
|  | 370.23 | 367.33 |
|  | 372.06 | 367.86 |
|  | 375.85 | 369.49 |
|  | 382.31 | 370.64 |
|  |  | 373.38 |
|  |  | 378.84 |

**Table 2.** Calculated electronic bandgaps of GaAs/GaP SLs of different periodicities.

|  | $n_{atoms}$ | Period (nm) | $E_g$ LDA (eV) | $E_g$ $G_0W_0$ (eV) |
|---|---|---|---|---|
| WZ GaAs bulk | 4 |  | 0.55 | 1.46[64] |
| WZ GaP bulk | 4 |  | 1.32 | 2.28[64] |
| GaAs/GaP SL$_5$ | 32 | 5.11 | 0.45 | 1.38* |
| GaAs/GaP SL$_6$ | 40 | 6.39 | 0.43 | 1.36* |
| GaAs/GaP SL$_{10}$ | 64 | 10.22 | 0.36 | 1.29* |

* estimates based on the average $G_0W_0$ corrections for bulk GaAs WZ and GaP WZ of Giorgi *et al.*[64]

AUTHOR INFORMATION

**Corresponding Author**

*E-mail: ilaria.zardo@unibas.ch

**Author Contributions**

I.Z. conceived the experiment. A.K.S., B.A., and J.T. performed the Raman measurements, which were analyzed by A.K.S. The samples were grown by O.A., V.Z., and L.S, while C. A. and J.T. transferred the wires and patterned the substrates. T.A. and R.R. performed the theoretical calculations. A.R.C. performed EDX analysis and TEM investigations. A.K.S., R.R. and I.Z. wrote the manuscript with contributions of all authors. All authors have given approval to the final version of the manuscript.

**Supporting Information**

- Contains further information on TEM analysis, Raman spectra in different polarization configuration, calculated phonon dispersion of SLs with different periodicity, eigen displacement of phonon modes in a SL unit cell, and analysis of BLS spectrum.

ACKNOWLEDGMENT

This project has received funding from the European Research Council (ERC) under the European Union's Horizon 2020 research and innovation program (grant agreement No 756365) and from the Swiss National Science Foundation grant (Grant No. 200021_184942). A.K.S. acknowledges financial support by the Georg H. Endress foundation. R.R. acknowledges financial support by the Ministerio de Economía, Industria y Competitividad (MINECO) under grant FEDER-MAT2017-90024-P and the Severo Ochoa Centres of Excellence Program under grant SEV-2015-0496 and by the Generalitat de Catalunya under grant no. 2017 SGR 1506. B.A. acknowledges support from the European Union's Horizon 2020 research and innovation programme under the Marie Skłodowska-Curie Grant Agreement No. 891443. A.R.C. acknowledges financial support from the NCCR SPIN, a National Centre of Competence (or Excellence) in Research, funded by the Swiss National Science Foundation (Grant No. 51NF40-180604).

REFERENCES

(1) Pop, E.; Sinha, S.; Goodson, K. E. Heat Generation and Transport in Nanometer-Scale Transistors. *Proceedings of the IEEE* **2006**, *94* (8), 1587–1601.
(2) Pop, E. Energy Dissipation and Transport in Nanoscale Devices. *Nano Res.* **2010**, *3* (3), 147–169.
(3) Yu, J.-K.; Mitrovic, S.; Tham, D.; Varghese, J.; Heath, J. R. Reduction of Thermal Conductivity in Phononic Nanomesh Structures. *Nat. Nanotechnol.* **2010**, *5* (10), 718–721.
(4) Moore, A. L.; Shi, L. Emerging Challenges and Materials for Thermal Management of Electronics. *Mater. Today* **2014**, *17* (4), 163–174.
(5) Williams, B. S. Terahertz Quantum-Cascade Lasers. *Nat. Photonics* **2007**, *1* (9), 517–525.
(6) Hochbaum, A. I.; Chen, R.; Delgado, R. D.; Liang, W.; Garnett, E. C.; Najarian, M.; Majumdar, A.; Yang, P. Enhanced Thermoelectric Performance of Rough Silicon Nanowires. *Nature* **2008**, *451* (7175), 163–167.
(7) Padture, N. P.; Gell, M.; Jordan, E. H. Thermal Barrier Coatings for Gas-Turbine Engine Applications. *Science* **2002**, *296* (5566), 280–284.
(8) Maldovan, M. Sound and Heat Revolutions in Phononics. *Nature* **2013**, *503* (7475), 209–217.
(9) Peierls, R. Zur kinetischen Theorie der Wärmeleitung in Kristallen. *Ann. Phys.* **1929**, 395(8), 1055-1101.
(10) Maldovan, M. Phonon Wave Interference and Thermal Bandgap Materials. *Nat. Mater.* **2015**, *14* (7), 667–674.
(11) Ravichandran, J.; Yadav, A. K.; Cheaito, R.; Rossen, P. B.; Soukiassian, A.; Suresha, S. J.; Duda, J. C.; Foley, B. M.; Lee, C.-H.; Zhu, Y.; Lichtenberger, A. W.; Moore, J. E.; Muller, D. A.; Schlom, D. G.; Hopkins, P. E.; Majumdar, A.; Ramesh, R.; Zurbuchen, M. A. Crossover from Incoherent to Coherent Phonon Scattering in Epitaxial Oxide Superlattices. *Nat. Mater.* **2014**, *13* (2), 168–172.
(12) Alaie, S.; Goettler, D. F.; Su, M.; Leseman, Z. C.; Reinke, C. M.; El-Kady, I. Thermal Transport in Phononic Crystals and the Observation of Coherent Phonon Scattering at Room Temperature. *Nat. Commun.* **2015**, *6* (1), 7228.
(13) Iskandar, A.; Gwiazda, A.; Huang, Y.; Kazan, M.; Bruyant, A.; Tabbal, M.; Lerondel, G. Modification of the Phonon Spectrum of Bulk Si through Surface Nanostructuring. *J. Appl. Phys.* **2016**, *120* (9), 095106.
(14) Yang, L.; Yang, N.; Li, B. Extreme Low Thermal Conductivity in Nanoscale 3D Si Phononic Crystal with Spherical Pores. *Nano Lett.* **2014**, *14* (4), 1734–1738.
(15) Jin, Y.; Pennec, Y.; Bonello, B.; Honarvar, H.; Dobrzynski, L.; Djafari-Rouhani, B.; Hussein, M. I. Physics of Surface Vibrational Resonances: Pillared Phononic Crystals, Metamaterials, and Metasurfaces. *Rep. Prog. Phys.* **2021**, *84* (8), 086502.
(16) Kushwaha, M. S.; Halevi, P.; Dobrzynski, L.; Djafari-Rouhani, B. Acoustic Band Structure of Periodic Elastic Composites. *Phys. Rev. Lett.* **1993**, *71* (13), 2022–2025.
(17) De Luca, M.; Fasolato, C.; Verheijen, M. A.; Ren, Y.; Swinkels, M. Y.; Kölling, S.; Bakkers, E. P. A. M.; Rurali, R.; Cartoixà, X.; Zardo, I. Phonon Engineering in Twinning Superlattice Nanowires. *Nano Lett.* **2019**, *19* (7), 4702–4711.
(18) Hoglund, E. R.; Bao, D.-L.; O'Hara, A.; Makarem, S.; Piontkowski, Z. T.; Matson, J. R.; Yadav, A. K.; Haislmaier, R. C.; Engel-Herbert, R.; Ihlefeld, J. F.; Ravichandran, J.; Ramesh, R.; Caldwell, J. D.; Beechem, T. E.; Tomko, J. A.; Hachtel, J. A.; Pantelides, S. T.; Hopkins,

P. E.; Howe, J. M. Emergent Interface Vibrational Structure of Oxide Superlattices. *Nature* **2022**, *601* (7894), 556–561.
(19) Li, D.; Wu, Y.; Fan, R.; Yang, P.; Majumdar, A. Thermal Conductivity of Si/SiGe Superlattice Nanowires. *Appl. Phys. Lett.* **2003**, *83* (15), 3186–3188.
(20) Luckyanova, M. N.; Garg, J.; Esfarjani, K.; Jandl, A.; Bulsara, M. T.; Schmidt, A. J.; Minnich, A. J.; Chen, S.; Dresselhaus, M. S.; Ren, Z.; Fitzgerald, E. A.; Chen, G. Coherent Phonon Heat Conduction in Superlattices. *Science* **2012**, *338* (6109), 936–939.
(21) Kitchin, M. R.; Shaw, M. J.; Corbin, E.; Hagon, J. P.; Jaros, M. Optical Properties of Imperfect Strained-Layer InAs/Ga$_{1-x}$In$_x$Sb/AlSb Superlattices with Infrared Applications. *Phys. Rev. B* **2000**, 61 (12), 8375–8381.
(22) Yao, T. Thermal Properties of AlAs/GaAs Superlattices. *Appl. Phys. Lett.* **1987**, *51* (22), 1798–1800.
(23) Hyldgaard, P.; Mahan, G. D. Phonon Superlattice Transport. *Phys. Rev. B* **1997**, *56* (17), 10754–10757.
(24) Tamura, S.; Tanaka, Y.; Maris, H. J. Phonon Group Velocity and Thermal Conduction in Superlattices. *Phys. Rev. B* **1999**, *60* (4), 2627–2630.
(25) Bugallo, D.; Langenberg, E.; Carbó-Argibay, E.; Varela Dominguez, N.; Fumega, A. O.; Pardo, V.; Lucas, I.; Morellón, L.; Rivadulla, F. Tuning Coherent-Phonon Heat Transport in LaCoO3/SrTiO3 Superlattices. *J. Phys. Chem. Lett.* **2021**, *12* (49), 11878–11885.
(26) Chen, X.-K.; Xie, Z.-X.; Zhou, W.-X.; Tang, L.-M.; Chen, K.-Q. Phonon Wave Interference in Graphene and Boron Nitride Superlattice. *Appl. Phys. Lett.* **2016**, *109* (2), 023101.
(27) Simkin, M. V.; Mahan, G. D. Minimum Thermal Conductivity of Superlattices. *Phys. Rev. Lett.* **2000**, *84* (5), 927–930.
(28) Gudiksen, M. S.; Lauhon, L. J.; Wang, J.; Smith, D. C.; Lieber, C. M. Growth of Nanowire Superlattice Structures for Nanoscale Photonics and Electronics. *Nature* **2002**, *415* (6872), 617–620.
(29) Paladugu, M.; Zou, J.; Guo, Y.-N.; Auchterlonie, G. J.; Joyce, H. J.; Gao, Q.; Hoe Tan, H.; Jagadish, C.; Kim, Y. Novel Growth Phenomena Observed in Axial InAs/GaAs Nanowire Heterostructures. *Small* **2007**, *3* (11), 1873–1877.
(30) Verheijen, M. A.; Immink, G.; de Smet, T.; Borgström, M. T.; Bakkers, E. P. A. M. Growth Kinetics of Heterostructured GaP−GaAs Nanowires. *J. Am. Chem. Soc.* **2006**, *128* (4), 1353–1359.
(31) Dick, K. A.; Thelander, C.; Samuelson, L.; Caroff, P. Crystal Phase Engineering in Single InAs Nanowires. *Nano Lett.* **2010**, *10* (9), 3494–3499.
(32) Caroff, P.; Dick, K. A.; Johansson, J.; Messing, M. E.; Deppert, K.; Samuelson, L. Controlled Polytypic and Twin-Plane Superlattices in III–V Nanowires. *Nat. Nanotechnol.* **2009**, *4* (1), 50–55.
(33) Johansson, J.; Dick, K. A. Recent Advances in Semiconductor Nanowire Heterostructures. *CrystEngComm* **2011**, *13* (24), 7175–7184.
(34) Glas, F. Critical Dimensions for the Plastic Relaxation of Strained Axial Heterostructures in Free-Standing Nanowires. *Phys. Rev. B* **2006**, *74* (12), 121302.
(35) Glas, F. Chapter Two - Strain in Nanowires and Nanowire Heterostructures. In *Semiconductors and Semimetals*; Morral, A. F. I., Dayeh, S. A., Jagadish, C., Eds.; Semiconductor Nanowires I; Elsevier **2015,** Vol. 93, 79–123.
(36) Ghukasyan, A.; LaPierre, R. Thermal Transport in Twinning Superlattice and Mixed-Phase GaAs Nanowires. *Nanoscale* **2022**, *14* (17), 6480–6487.

(37) López-Güell, K.; Forrer, N.; Cartoixà, X.; Zardo, I.; Rurali, R. Phonon Transport in GaAs and InAs Twinning Superlattices. *J. Phys. Chem. C* **2022**, *126* (39), 16851–16858.
(38) Peri, L.; Prete, D.; Demontis, V.; Zannier, V.; Rossi, F.; Sorba, L.; Beltram, F.; Rossella, F. Giant Reduction of Thermal Conductivity and Enhancement of Thermoelectric Performance in Twinning Superlattice InAsSb Nanowires. *Nano Energy* **2022**, *103*, 107700.
(39) Steiner, M.A.; France, R.M.; Buencuerpo, J.; Geisz, J.F.; Nielsen, M.P.; Pusch, A.; Olavarria, W.J.; Young, M.; Ekins-Daukes, N.J. High efficiency inverted GaAs and GaInP/GaAs solar cells with strain-balanced GaInAs/GaAsP quantum well*s, Adv. Energy Mater* **2021**, 11 (2021), 2002874.
(40) Li, Z.; Kim, T.; H., Han, S. Y.; Yun, Y.-J.; Jeong, S.; Jo, B.; Ok, S. A.; Yim, W.; Lee, S. H.; Kim, K.; Moon, S.; Park, J.-Y.; Ahn, T. K.; Shin, H.; Lee, J.; Park, H. J.; Wide-Bandgap Perovskite/Gallium Arsenide Tandem Solar Cells. Adv. Energy Mater. *2020*, 10(6), 1903085.
(41) Siao, H.-Y.; Bunk, R. J.; Woodall, J. M. Gallium Phosphide Solar Cell Structures with Improved Quantum Efficiencies. *J. Electron. Mater.* **2020**, *49* (6), 3435–3440.
(42) Melli, M.; West, M.; Hickman, S.; Dhuey, S.; Lin, D.; Khorasaninejad, M.; Chang, C.; Jolly, S.; Tae, H.; Poliakov, E.; St. Hilaire, P.; Cabrini, S.; Peroz, C.; Klug, M. Gallium Phosphide Optical Metasurfaces for Visible Light Applications. *Sci Rep* **2020,** 10 (1), 20694.
(43) Li, Q.; Fang, S; Liu, S; Xu, L; Xu, L; Yang, C; Yang, J; Shi, B; Ma, J; Yang, J; Quhe, R; Lu, J. Performance Limit of Ultrathin GaAs Transistors. *ACS Appl Mater Interfaces.* **2022,** 14(20), 23597–23609.
(44) Arif, O.; Zannier, V.; Rossi, F.; Matteis, D. D.; Kress, K.; Luca, M. D.; Zardo, I.; Sorba, L. GaAs/GaP Superlattice Nanowires: Growth, Vibrational and Optical Properties. *Nanoscale* **2023**, *15* (3), 1145–1153.
(45) Gupta, R.; Xiong, Q.; Mahan, G. D.; Eklund, P. C. Surface Optical Phonons in Gallium Phosphide Nanowires. *Nano Lett.* **2003**, *3* (12), 1745–1750.
(46) Mahan, G. D.; Gupta, R.; Xiong, Q.; Adu, C. K.; Eklund, P. C. Optical Phonons in Polar Semiconductor Nanowires. *Phys. Rev. B* **2003**, *68* (7), 073402.
(47) Mooradian, A.; Wright, G. B. First Order Raman Effect in III–V Compounds. *Solid State Commun.* **1966**, *4* (9), 431–434.
(48) Ramsteiner, M.; Brandt, O.; Kusch, P.; Breuer, S.; Reich, S.; Geelhaar, L. Quenching of the E2 Phonon Line in the Raman Spectra of Wurtzite GaAs Nanowires Caused by the Dielectric Polarization Contrast. *Appl. Phys. Lett.* **2013**, *103* (4), 043121.
(49) Widulle, F.; Ruf, T.; Göbel, A.; Schönherr, E.; Cardona, M. Raman Study of the Anomalous TO Phonon Structure in GaP with Controlled Isotopic Composition. *Phys. Rev. Lett.* **1999**, *82* (26), 5281–5284.
(50) Zardo, I.; Conesa-Boj, S.; Peiro, F.; Morante, J. R.; Arbiol, J.; Uccelli, E.; Abstreiter, G.; Fontcuberta i Morral, A. Raman Spectroscopy of Wurtzite and Zinc-Blende GaAs Nanowires: Polarization Dependence, Selection Rules, and Strain Effects. *Phys. Rev. B* **2009**, *80* (24), 245324.
(51) De Luca, M.; Zardo, I. Semiconductor Nanowires: Raman Spectroscopy Studies. In Raman Spectroscopy and Applications; Maaz, K., Ed.; *InTech*, **2017**, 81-101.
(52) Hayashi, S.; Kanamori, H. Raman Scattering from the Surface Phonon Mode in GaP Microcrystals. *Phys. Rev. B* **1982**, *26* (12), 7079–7082.

(53) Jia, X.; Lin, Z.; Zhang, T.; Puthen-Veettil, B.; Yang, T.; Nomoto, K.; Ding, J.; Conibeer, G.; Perez-Wurfl, I. Accurate Analysis of the Size Distribution and Crystallinity of Boron Doped Si Nanocrystals via Raman and PL Spectra. *RSC Adv.* **2017**, *7* (54), 34244–34250.

(54) Jusserand, B.; Cardona, M. Raman Spectroscopy of Vibrations in Superlattices. In *Light Scattering in Solids V: Superlattices and Other Microstructures*; Cardona, M., Güntherodt, G., Eds.; Topics in Applied Physics; Springer: Berlin, Heidelberg, **1989,** 49–152.

(55) Scarponi, F.; Mattana, S.; Corezzi, S.; Caponi, S.; Comez, L.; Sassi, P.; Morresi, A.; Paolantoni, M.; Urbanelli, L.; Emiliani, C.; Roscini, L.; Corte, L.; Cardinali, G.; Palombo, F.; Sandercock, J. R.; Fioretto, D. High-Performance Versatile Setup for Simultaneous Brillouin-Raman Microspectroscopy. *Phys. Rev. X* **2017**, *7* (3), 031015.

(56) Sandercock, J. R. Trends in Brillouin Scattering: Studies of Opaque Materials, Supported Films, and Central Modes. In *Light Scattering in Solids III: Recent Results*; Cardona, M., Güntherodt, G., Eds.; Topics in Applied Physics; Springer: Berlin, Heidelberg **1982,** 173–206.

(57) Kargar, F.; Debnath, B.; Kakko, J.-P.; Säynätjoki, A.; Lipsanen, H.; Nika, D. L.; Lake, R. K.; Balandin, A. A. Direct Observation of Confined Acoustic Phonon Polarization Branches in Free-Standing Semiconductor Nanowires. *Nat. Commun.* **2016**, *7* (1), 13400.

(58) Mutti, P.; Bottani, C. E.; Ghislotti, G.; Beghi, M.; Briggs, G. A. D.; Sandercock, J. R. Surface Brillouin Scattering—Extending Surface Wave Measurements to 20 GHz. In *Advances in Acoustic Microscopy*; Briggs, A., Ed.; Advances in Acoustic Microscopy; Springer US: Boston, MA, 1995; pp 249–300.

(59) Bortolani, V.; Nizzoli, F.; Santoro, G. Surface Density of Acoustic Phonons in GaAs. *Phys. Rev. Lett.* **1978**, *41* (1), 39–42.

(60) Grimsditch, M.; Cardona, M.; Calleja, J. M.; Meseguer, F. Resonance in the Raman Scattering of CaF2, SrF2, BaF2 and Diamond. *J. Raman Spectrosc.* **1981**, *10* (1), 77–81.

(61) Klar, P.; Lidorikis, E.; Eckmann, A.; Verzhbitskiy, I. A.; Ferrari, A. C.; Casiraghi, C. Raman Scattering Efficiency of Graphene. *Phys. Rev. B* **2013**, *87* (20), 205435.

(62) Panda, J. K.; Roy, A.; Gemmi, M.; Husanu, E.; Li, A.; Ercolani, D.; Sorba, L. Electronic Band Structure of Wurtzite GaP Nanowires via Temperature Dependent Resonance Raman Spectroscopy. *Appl. Phys. Lett.* **2013**, *103* (2), 023108.

(63) Kresse, G.; Furthmüller, J. Efficient Iterative Schemes for Ab Initio Total-Energy Calculations Using a Plane-Wave Basis Set. *Phys. Rev. B* **1996**, *54* (16), 11169–11186.

(64) Giorgi, G.; Amato, M.; Ossicini, S.; Cartoixà, X.; Canadell, E.; Rurali, R. Doping of III–V Arsenide and Phosphide Wurtzite Semiconductors. *J. Phys. Chem. C* **2020**, *124* (49), 27203–27212.

(65) Kargar, F.; Balandin, A. A. Advances in Brillouin–Mandelstam Light-Scattering Spectroscopy. *Nat. Photonics* **2021**, *15* (10), 720–731.

(66) Baroni, S.; de Gironcoli, S.; Dal Corso, A.; Giannozzi, P. Phonons and Related Crystal Properties from Density-Functional Perturbation Theory. *Rev. Mod. Phys.* **2001**, *73* (2), 515–562.

(67) Gonze, X.; Amadon, B.; Anglade, P.-M.; Beuken, J.-M.; Bottin, F.; Boulanger, P.; Bruneval, F.; Caliste, D.; Caracas, R.; Côté, M.; Deutsch, T.; Genovese, L.; Ghosez, Ph.; Giantomassi, M.; Goedecker, S.; Hamann, D. R.; Hermet, P.; Jollet, F.; Jomard, G.; Leroux, S.; Mancini, M.; Mazevet, S.; Oliveira, M. J. T.; Onida, G.; Pouillon, Y.; Rangel, T.; Rignanese, G.-M.; Sangalli, D.; Shaltaf, R.; Torrent, M.; Verstraete, M. J.; Zerah, G.; Zwanziger, J. W. ABINIT:

First-Principles Approach to Material and Nanosystem Properties. *Comput. Phys. Commun.* **2009**, *180* (12), 2582–2615.
(68) Veithen, M.; Gonze, X.; Ghosez, Ph. Nonlinear Optical Susceptibilities, Raman Efficiencies, and Electro-Optic Tensors from First-Principles Density Functional Perturbation Theory. *Phys. Rev. B* **2005**, *71* (12), 125107.
(69) Gonze, X.; Vigneron, J.-P. Density-Functional Approach to Nonlinear-Response Coefficients of Solids. *Phys. Rev. B* **1989**, *39* (18), 13120–13128.

# Supporting Information

**Supporting Information 1: HR-TEM of GaAs/GaP SL nanowires**

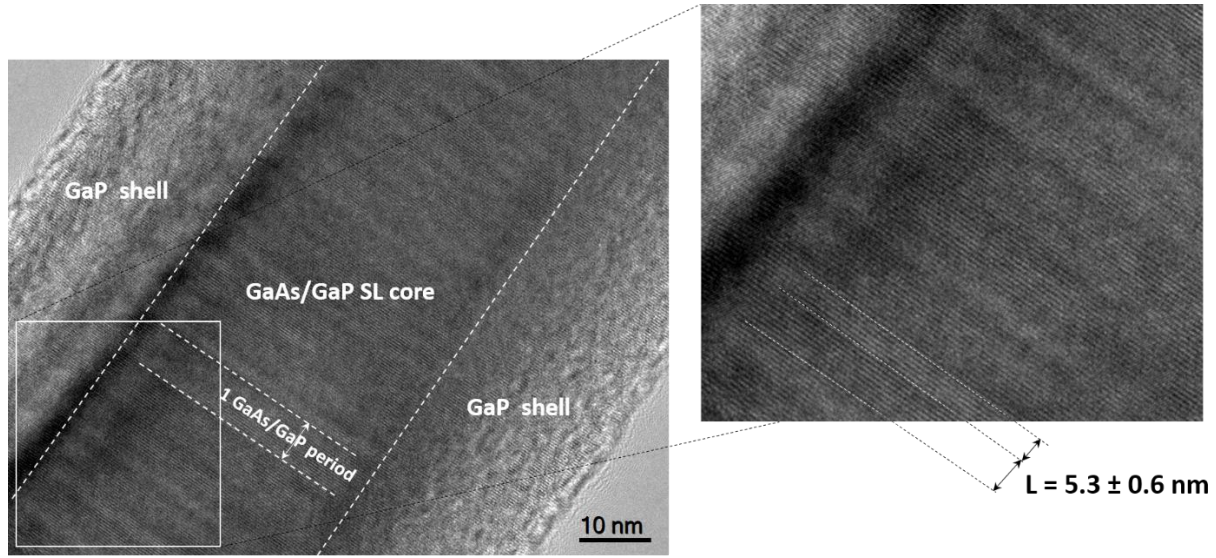

Figure S1. HR-TEM of a GaAs/GaP SL NW with nominal period of 4.8 nm. The inset zooms in on the SL segment.

In Figure S1, we show the HR-TEM of a GaAs/GaP SL NW with 4.8 nm periodicity. We used the software DigitalMicrograph$^{TM}$ on the HR-TEM and measured a period of 5.3 ± 0.6 nm, which is in good agreement with the nominal period (4.8 nm) determined through the growth protocol.

**Supporting information 2: Polarization dependent Raman scattering experiments**

In figure S2(a), we present results on the μ-Raman measurements in the $\bar{x}(y,y)x$ scattering configuration on the reference NW. For comparison we also present the theoretically calculated spectra of WZ GaP in figure S2(b).

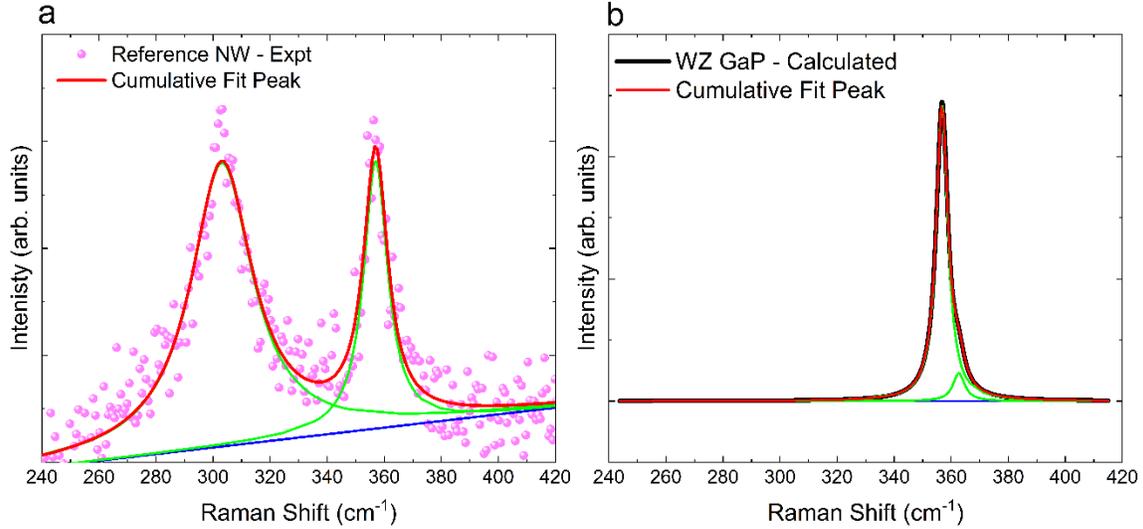

Figure S2(a) Polarized μ-Raman spectrum on a reference NW collected in the $\bar{x}(y,y)x$ scattering configuration. The experimental data are shown in pink spheres, light green curves show the individual Lorentzian fitting, the red curve shows the cumulative fitting and dark blue line is the baseline. (b) Theoretically calculated spectrum of WZ GaP in black curve, with the full width at half maximum fixed at 5 cm$^{-1}$. The green curves show the individual Lorentzian fitting and the red curve shows the cumulative fitting. The dark blue line is the baseline.

In Figure S2(a), we see a peak at 356.9 cm$^{-1}$ which can be attributed to the $E_2^H$ of WZ GaP[1]. There is another peak at 303 cm$^{-1}$, which is due to the silicon substrate[2]. In Figure S2(b), we present the calculated spectrum of WZ GaP. The spectrum has two peaks at 356.8 cm$^{-1}$ and at 362.7 cm$^{-1}$, which can be attributed to the $E_2^H$ and TO of GaP, respectively. The presence of $E_2^H$ mode confirms the WZ crystal phase of the NWs. In the experimental spectrum, in the $\bar{x}(y,y)x$ configuration, the resolution of the TO mode difficult which is why it is not used in the fitting in Figure S2 (a).

In Figure S3 (a), we present the results on the μ-Raman measurements in the $\bar{x}(y,y)x$ and $\bar{x}(z,z)x$ on SL NW with a period of 10 nm. In the $\bar{x}(z,z)x$ scattering configuration, the most intense peak in the GaAs-like phonon modes region is at 276.9 cm$^{-1}$ while in the GaP-like phonon modes region there is an intense peak at 357.6 cm$^{-1}$. In the $\bar{x}(y,y)x$ scattering configuration, the most intense peak in the GaAs-like phonon modes region is at 269.4 cm$^{-1}$ while in the GaP-like phonon modes region there is an intense peak at 351.6 cm-1. The overall intensity of the spectrum decreases in the $\bar{x}(y,y)x$ as compared to $\bar{x}(z,z)x$ configuration by a factor of about 2, possibly due also to the dielectric mismatch effect.[3] In Figure S3 (b), we show the calculated Raman spectrum of GaAs/GaP SL with L=6.39 nm in both the polarization configuration. The intensities are normalized for ease of comparison.

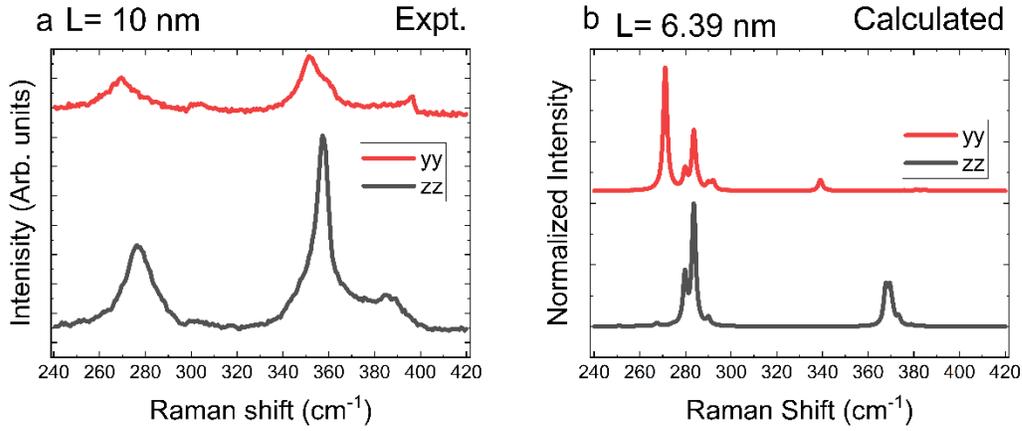

Figure S3. (a) Polarization resolved μ-Raman spectra of a 10 nm SL NW. The red curve shows the data collected in the $\bar{x}(y,y)x$ configuration and the black curve shows the data collected in the $\bar{x}(z,z)x$; (b) Calculated Raman spectra of a GaAs/GaP SL of L=6.39 nm. The intensities are normalized. The red curve shows the $\bar{x}(y,y)x$ configuration and black curves show the $\bar{x}(z,z)x$.

In Table S1 we list the computed Γ-point frequencies, setting a threshold intensity indicative of their experimental detectivity.

**Table S1.** The calculated frequencies of Raman active phonon modes for a GaAs/GaP SL structure with a period of 5.12 nm. Phonon modes with a Raman intensity more than 100 times smaller than the one of the most intense modes were considered non-detectable.

| Frequency (cm$^{-1}$) (Calculated for 5.12 nm period) | Ratio of Intensity/Highest Intensity (Calculated for 5.12 nm period) | Frequency (cm$^{-1}$) (Experimental for 4.8 nm period) | Ratio of Intensity/Highest Intensity (Experimental for 4.8 nm period) |
|---|---|---|---|
| 245.55 | 5E-03 | 261.3 | 8E-02 |
| 254.87 | 1E-02 | 276.4 | 9E-01 |
| 264.73 | 1E-02 | 287.6 | 5E-02 |
| 273.24 | 9E-02 | | |
| 281.13 | 7E-02 | | |
| 282.45 | 1 | | |
| 288.12 | 1E-01 | | |
| 367.25 | 2E-01 | 352.8 | 7E-01 |
| 368.09 | 1E-01 | 358.3 | 1 |
| 370.23 | 2E-01 | 368.1 | 8E -02 |
| 372.06 | 2E-02 | 376.7 | 7E-02 |
| 375.85 | 3E-02 | 388.2 | 1E-01 |
| 382.31 | 3E-03 | 391.98 | 2E-01 |
| | | 394.7 | 1E-01 |

**Supporting information 3: Phonon dispersion of GaAs/GaP SL with a period of 6.39 nm**

The phonon dispersion of the GaP/GaAs SL with period 6.39 nm obtained through *ab initio* calculations is displayed in Figure S4. The increased number of phonon modes at the Γ-point can be appreciated.

*Ab initio* calculations provide us also with the relaxed lattice parameters of the constituents of the superlattice. We obtain lattice parameters $a = 3.947$ and $c = 6.509$ Å for wurtzite GaAs and $a = 3.796$ and $c = 6.262$ Å for wurtzite GaP. Since in all the SLs investigated the length of GaAs and GaP segments is very close and the elastic constants of the constituent materials are similar, the lattice parameter, $a$, of all the cases studied is roughly the average of the bulk cases, i.e. 3.871 Å. The $c$ lattice parameter, on the other hand, must accommodate the period of each SL and is 12.775, 38.328, 51.103, and 63.880 Å. Neglecting the chemical identity in all SLs, we would obtain an "effective" $c$ lattice parameter of 6.388 Å that, again, is essentially the average of the values of wurtzite GaAs and wurtzite GaP.

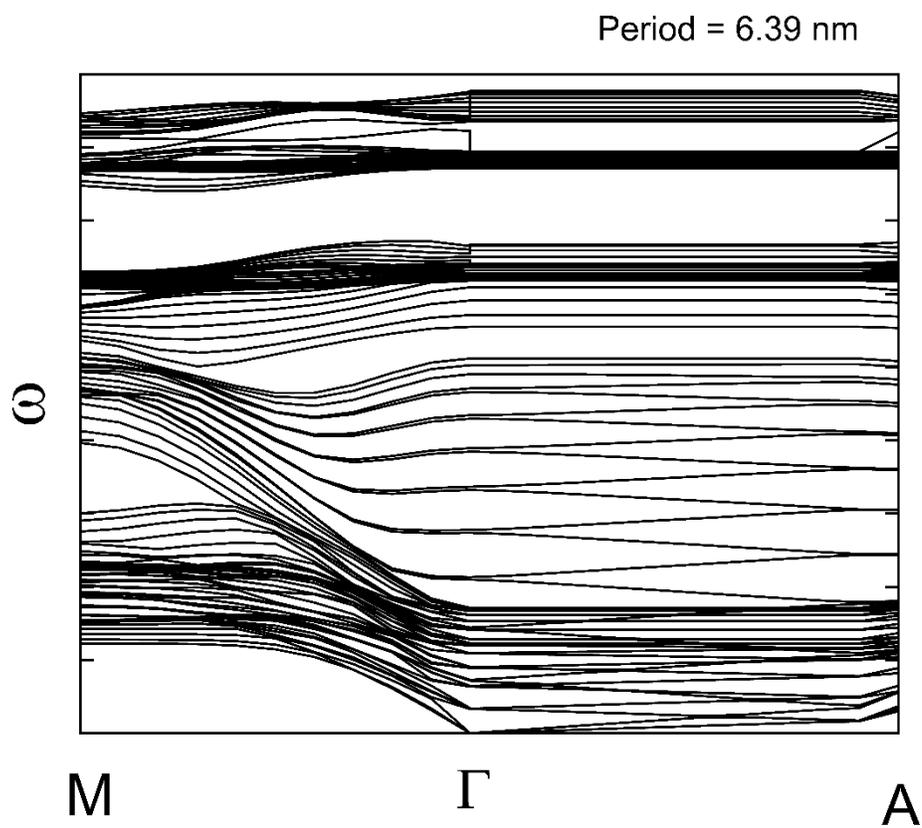

Figure S*4*. Calculated phonon dispersion of a GaAs/GaP SL NW *with* period 6.39 nm.

Table S2 – Calculated acoustic phonon mode frequencies with frequencies less than 1 THz for peaks corresponding to q value close to 85 μm$^{-1}$ from the dispersion relation in Figure S3 for SL with periodicity of 6.3 nm.

| Period | Frequency (GHz) |
|---|---|
| 6.3 nm | 33 |
| | 74 |
| | 444 |
| | 476 |
| | 814 |
| | 835 |
| | 877 |

**Supporting information 4: Eigen displacement of phonon modes involving different atoms in the SL unit cell**

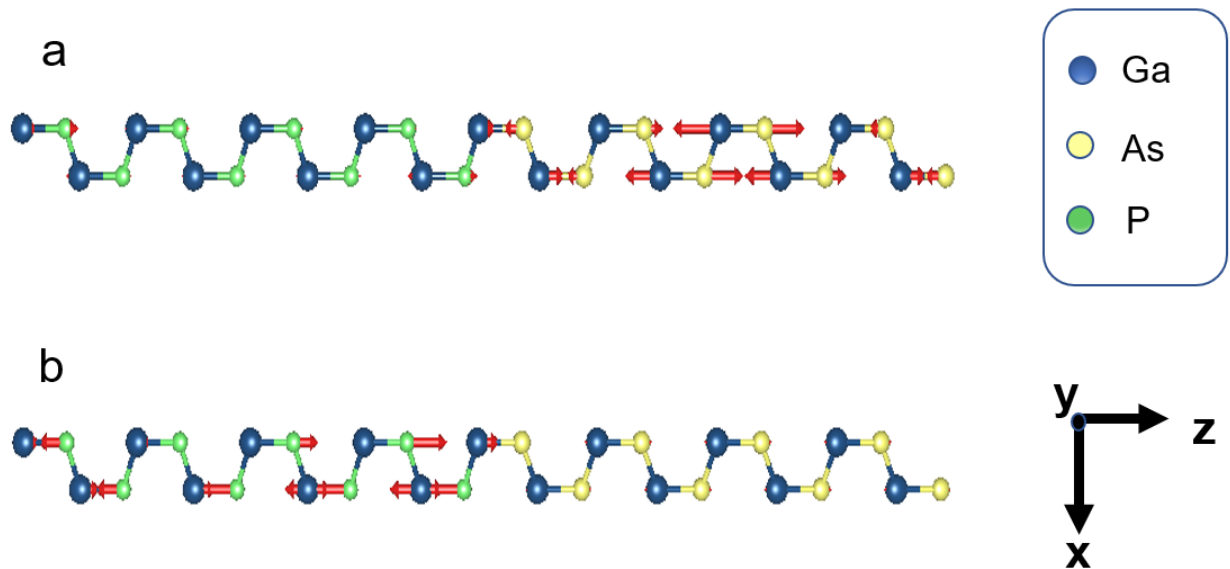

**Figure S5.** The eigen-displacement of atoms for phonon mode at (a) 288.12 cm$^{-1}$ and (b) 367.25 cm$^{-1}$. The different colored spheres represent the constituent atoms with blue as Ga, yellow as As, and green as P.

In Figure S5, show the schematic for two phonon modes at 288.1 cm$^{-1}$ and 367.2 cm$^{-1}$, respectively, selected from Table S1. The Ga atoms are represented by blue spheres, As by yellow spheres, and P with be green spheres.

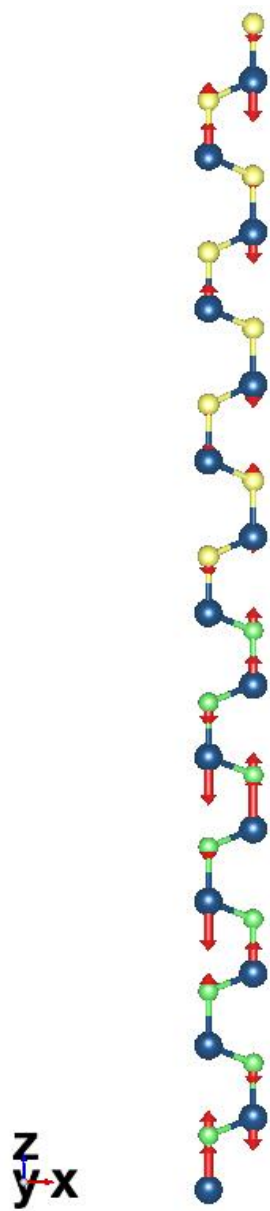

Figure S6. The Eigen displacement of atoms for phonon mode at 218 cm$^{-1}$. The different coloured spheres represent the constituent atoms with blue as Ga, yellow as As, and green as P.

In Figure S6, we show the eigen displacement of the atoms of the Raman inactive phonon mode at 218 cm$^{-1}$, whose vibrations involve all atoms in the SL unit cell. The Ga atoms are represented by blue spheres, As by yellow spheres, and P will be green spheres.

**Supporting information 4: Analysis of Brillouin light scattering interferometry**

In Figure S7, we present an example of data analysis of BLS spectrum. The data are fitted using deconvoluted Lorentzian curves as shown as green solid lines in the Figure. The cumulative fit is shown as red solid line.

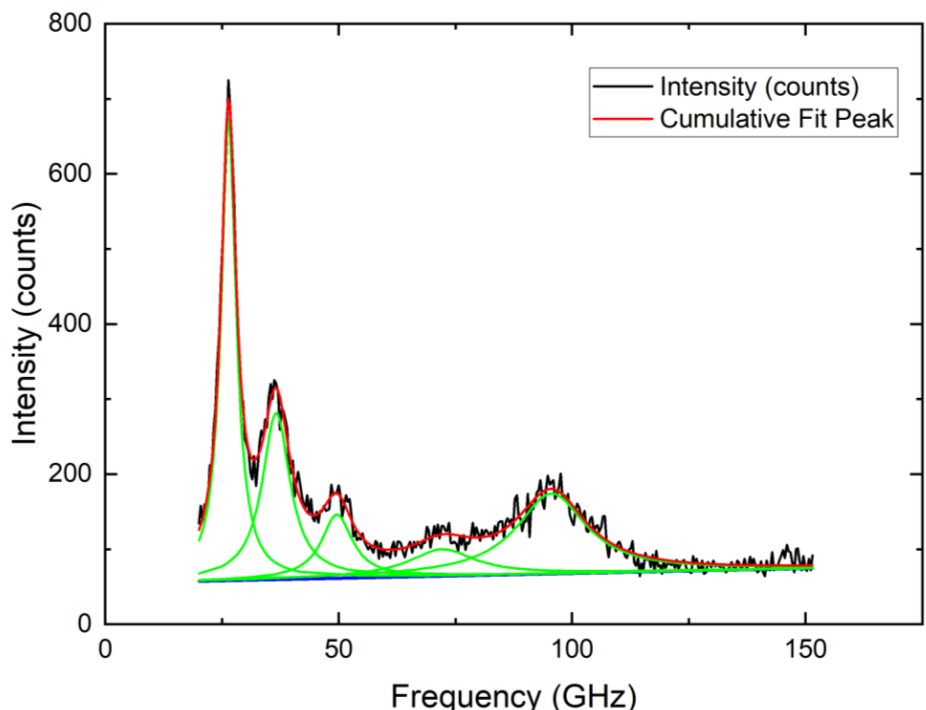

Figure S7. An example of BLS spectrum analysis. An example of deconvolution for a SL NW with period 6 nm. To analyse the data, we used peak deconvolution with several Lorentzian functions, shown in green curves. The black curve is the expt. Data and the red curve is cumulative fit.

References

(1) Gupta, R.; Xiong, Q.; Mahan, G. D.; Eklund, P. C. Surface Optical Phonons in Gallium Phosphide Nanowires. *Nano Lett.* **2003**, *3* (12), 1745–1750.
(2) Jia, X.; Lin, Z.; Zhang, T.; Puthen-Veettil, B.; Yang, T.; Nomoto, K.; Ding, J.; Conibeer, G.; Perez-Wurfl, I. Accurate Analysis of the Size Distribution and Crystallinity of Boron Doped Si Nanocrystals via Raman and PL Spectra. *RSC Adv.* **2017**, *7* (54), 34244–34250.
(3) Zardo, I.; Conesa-Boj, S.; Peiro, F.; Morante, J. R.; Arbiol, J.; Uccelli, E.; Abstreiter, G.; Fontcuberta i Morral, A. Raman Spectroscopy of Wurtzite and Zinc-Blende GaAs

Nanowires: Polarization Dependence, Selection Rules, and Strain Effects. *Phys. Rev. B* **2009**, *80* (24), 245324.